\documentclass[12pt]{article} 
\usepackage{epsfig}
\def\ical{{\cal I}}
\def\lcal{{\cal L}}

\def\inv#1{{1\over #1}}
\def\half{\inv2}

\def\gesim{\,{\raise-3pt\hbox{$\sim$}}\!\!\!\!\!{\raise2pt\hbox{$>$}}\,}
\def\lesim{\,{\raise-3pt\hbox{$\sim$}}\!\!\!\!\!{\raise2pt\hbox{$<$}}\,}
\def\ab{{\alpha,\beta}}
\def\sgn{{\,\rm Sgn}}

\newcommand{\nc}{\newcommand}
\nc{\beq}{\begin{equation}}  \nc{\eeq}{\end{equation}}
\nc{\bea}{\begin{eqnarray}}  \nc{\eea}{\end{eqnarray}}
\nc{\baa}{\begin{array}}     \nc{\eaa}{\end{array}}
\nc{\bit}{\begin{itemize}}   \nc{\eit}{\end{itemize}}
\nc{\ben}{\begin{enumerate}} \nc{\een}{\end{enumerate}}
\nc{\bce}{\begin{center}}    \nc{\ece}{\end{center}}
\def\ie{\hbox{\it i.e.}}

\def\eg{\hbox{\it e.g.}}

\def\etal{\hbox{\it et al.}}

\begin{document}

\begin{titlepage}

\title{\vskip-1in \hfill{{\small MCTP-02-47}} \vskip.5in
Interacting Quantum Field Theory\\ in de~Sitter Vacua}
\author{Martin B. Einhorn\footnote{\texttt{meinhorn@umich.edu}}\ \ and 
Finn Larsen\footnote{\texttt{larsenf@umich.edu}} \\
\centerline{\it\normalsize Michigan Center for Theoretical Physics,
Randall Laboratory of Physics} \\  
\centerline{\it\normalsize The University of Michigan, Ann Arbor, MI 48109-1120} }

\date{September 19, 2002}


\maketitle

\baselineskip=15pt

\begin{abstract}
We discuss interacting quantum field theory in de~Sitter space 
and argue that the Mottola-Allen vacuum ambiguity is an artifact of free field
theory. The nature of the nonthermality of the MA-vacua is also 
clarified. We propose analyticity of correlation functions as a fundamental 
requirement of quantum field theory in curved spacetimes. In de~Sitter space,  
this principle determines the vacuum unambiguously and
facilitates the systematic development of perturbation theory.
\end{abstract}

\thispagestyle{empty}
\end{titlepage}

\setcounter{page}{1}

\section{Introduction}
The quantum theory of de~Sitter space is important because it plays
a central role in cosmology, particularly in the generation of
cosmic structure from inflation and in the puzzles 
surrounding the cosmological constant. More generally, de~Sitter space is 
useful for testing theoretical ideas because, as a maximally 
symmetric space, it is tightly constrained. Eventually, 
we want to understand the full quantum gravity of de~Sitter space, but many 
interesting questions appear already at the level of quantum field theory 
(QFT) in a fixed de~Sitter background. In this paper we study the vacuum 
structure of the theory and develop the features of interacting QFT 
needed for this purpose.

The starting point of QFT in any background is a mode expansion
\begin{equation}
\phi(X) = \sum_{n} [ a_{n} u_{n}(X) + a_{n}^{\dagger}u_{n}^{*}(X)]~,
\label{modeexp}
\end{equation}
and the corresponding specification of the vacuum 
\begin{equation}
a_{n}|{\rm vac}\rangle = 0 ~.
\label{vacdef}
\end{equation}
The modes {$u_{n}$} must be chosen so that the corresponding vacuum respects the 
symmetries of the theory. In Minkowski space this principle determines the
modes 
completely and, accordingly, there is a unique Poincar\'{e} invariant vacuum. 
In curved spacetime, symmetries do not in general determine the vacuum state 
completely. Indeed, in de~Sitter space, symmetries identify the vacuum 
only modulo a two parameter ambiguity, 
corresponding to a family of distinct de~Sitter 
invariant vacua. The existence of this ambiguity was first emphasized
by Mottola\cite{Mottola:ar} and Allen\cite{Allen:ux}.

One vacuum is almost universally taken as the starting point for QFT 
in de~Sitter space. We will refer to this vacuum as the ``Euclidean vacuum''
and reserve the term ``MA-vacua'' for the alternative, nonstandard vacua. 
The Euclidean vacuum is singled out by several features:
\begin{itemize}
\item
The correlation functions can be obtained by continuation from the 
Euclidean de~Sitter space, {\it i.e.} a sphere. This is the origin
of the terminology we employ.
\item
It coincides with the adiabatic vacuum in the FRW coordinates customarily 
employed in cosmology. This facilitates a consistent particle interpretation 
of the theory. In cosmology the Euclidean vacuum is often referred to as the 
Bunch-Davies vacuum\cite{Bunch:yq}.
\item
The correlations of the field are experienced as precisely thermal
by an Unruh detector (a comoving detector linearly 
coupled to the quantum field). 
\item
The $2$-point correlation functions reduce to the standard Minkowski  
propagators when the de~Sitter radius is taken to infinity. 
\end{itemize}
These properties are desirable both technically and conceptually, but
they do not by themselves identify the Euclidean vacuum as the ``right'' 
vacuum. It could be that the MA-vacua simply have different
properties, and that the question of which vacuum is appropriate 
depends on additional physical input such as boundary conditions,
or even observational data. This is the point of view taken in much 
recent work\cite{SSV,BMS,SV,danielsson,transplanck,Balasubramanian:2002zh}.

Previous discussions have been at the level of free field theory in fixed background. In most
contexts, such as effective field theory, what we are really interested in is
the weakly interacting theory; so it is important that higher order interactions
can be included, at least in principle. In this paper we argue that this is
possible {\it only} for the Euclidean vacuum. The problems we encounter in the
general case are particularly sharp for the loop amplitudes. For these the issue
{\it is not} that amplitudes take values we deem physically unreasonable;
rather, {\it they are ill-defined in the MA-vacua}, a much worse problem. Tree
level amplitudes are also problematic even though they are mathematically well-defined: 
they have unusual, and most likely physically unacceptable,
singularities related to antipodal events which, as we discuss, cannot be hidden
behind an event horizon. We conclude that {\it the MA-vacua are artifacts 
of the free field limit}. Of course, we cannot actually {\it prove} that no 
definition of the MA-vacua exists at the interacting level. What
we argue is that it is known how
to include interactions in the Euclidean vacuum, and, whenever the propagator is
not the boundary value of an analytic function, as in
these MA-vacua, this prescription does not generalize.  Thus, at the very least,
more work is needed to establish the viability  of the MA-vacua.

Relativistic QFT is a tight structure. In addition to symmetries, the
interacting theory is constrained by analyticity properties. The principle we
need is
\begin{itemize}
\item
Correlation functions are boundary values of analytical functions
(in the sense of distributions).
\end{itemize}
This principle singles out the Euclidean vacuum. Following Bros \etal\cite{Bros:1990cu,Bros:dn,Bros:1995js,Bros:1998ik}, it allows one to construct
an interacting theory for de~Sitter space that closely mimics QFT in Minkowski space. 
For example, it ensures the K\"all\'en-Lehman representation for the two-point Green's
function with a positive spectral density. The non-existence of an S-matrix, a
major confusion clouding the quantum theory of de~Sitter space, can be
circumvented, at least for our purposes, by considering correlation functions
directly. It is the correlation functions that are observables, measured by
Unruh detectors, and it is the correlation functions that satisfy strong
analyticity properties. These may be important lessons for formulating the
quantum theory of de~Sitter space.

One of the motivations for studying the MA-vacua is their possible applications
to cosmology. According to the inflationary paradigm, all structure in the
universe ultimately originated from the fluctuations of a scalar field in a
de~Sitter background. It has been proposed that physics at extremely high
``trans-Planckian'' energies determines which vacuum is appropriate for this
scalar field and thus, using the MA-vacua as interlocutor, the cosmic structure
could contain data pertaining to such energies\cite{danielsson,transplanck}. In
this regard our results are unfortunately negative: they indicate that this
possibility is illusory, at least in its simplest form.

The remainder of the paper is organized as follows. In section 2 we review the
classical de~Sitter geometry, QFT in the de~Sitter background, and the MA-vacua
in the free theory. In section 3 we first discuss some general features of
interacting QFT in curved spacetimes and singularities of amplitudes. Then we
exhibit the problems with the MA-vacua, considering in turn tree level
amplitudes and loops. In section 4 we discuss the nature of the non-thermality
of the MA-vacua, and examine some of their difficulties from this point of view. 
Finally, in section 5, we discuss implications of our results as well as
future research directions.

\section{Quantum Field Theory in de~Sitter Space}
The purpose of this section is to review properties of the de~Sitter
geometry and QFT in the de~Sitter background with
special emphasis on the MA-vacua. The primary reference
is \cite{Allen:ux}.

\subsection{Classical Geometry}
The d-dimensional de~Sitter space ($dS_{d}$) can be represented as 
the hyperboloid
\begin{equation}
X^{2}=X^{2}_{0} - \vec{X}^{2} = -l^{2}~,
\label{dsconstraint}
\end{equation}
in $d+1$ dimensional Minkowski space 
\begin{equation}
X^M=(X^{0},X^{1},\ldots,X^{d})~,
\end{equation}
with signature $(+,-,\ldots,-)$. De~Sitter space is maximally 
symmetric with isometry group $SO(1,d-1)$; and it is a solution to pure 
gravity with cosmological constant 
$
\Lambda = {1\over 2}(d-1)(d-2)l^{-2}~.
$
We take $l^{2}=1$ in the following.

The $SO(1,d)$ invariant distance between different points $X$ and
$Y$ is 
\begin{equation}
Z=-X\cdot Y = \vec{X}\cdot\vec{Y} - X^{0}Y^{0}~.
\end{equation}
The $X$ and $Y$ are timelike separated if $Z>1$ and spacelike 
separated for $Z<1$. For $Z<-1,$ no geodesic exists that join $X$ 
and $Y$, even though the spacetime is geodesically complete.

The constraint equation defining de~Sitter space (\ref{dsconstraint}) is
invariant under $X\to -X$; so for any $X$ in de~Sitter space, the antipodal
event $X_{A}=-X$ is also in de~Sitter space. The invariant distance between an
event and its antipodal is $Z=-X\cdot X_{A}=-1$. This means their future (past)
lightcones cross only in the asymptotic future (past). Antipodal events cannot
communicate; they never have, and they never will.

The embedding coordinates $X^M$, and the invariant distance $Z$, are the most
convenient for our purposes. Various explicit coordinates, solving the embedding
relations, are more familiar in other contexts. For their relation to embedding
coordinates, see the recent review \cite{SSV}.

The invariant distance $Z=Z(X,Y)$ is symmetric in its two arguments; so it does
not distinguish between the future and past lightcone. For such purposes we
introduce $\sgn(X,Y)$; it is equal to $+1~(-1)$ when $X$ is in or on the forward
(backward) lightcone of $Y$, and equal to $0$ for spacelike directions. The
$\sgn(X,Y)$ is invariant under de~Sitter symmetries continuously connected to
the identity, but it changes sign under time reversal.

\subsection{Quantum Fields in de~Sitter Space}
We want to study quantum field theory in the de~Sitter background. The
simplest example is a free scalar field 
\begin{equation}
{\cal L} = {1\over 2}[ \partial_\mu \phi\partial^{\mu}\phi - m^{2}\phi^{2}
-\xi {\cal R}\phi^{2}]~.
\end{equation}
Since the scalar curvature ${\cal R}={\rm const}=d(d-1)$ in de~Sitter space,
we can absorb the coupling $\xi$ into an effective mass 
$\tilde{m}^{2}=m^{2}+\xi {\cal R}$ and henceforth consider minimal coupling
$\xi=0$ without loss of generality. The starting point for quantization 
of the field theory is a complete set $\{ u_{n} \}$ of properly normalized 
solutions to the Klein-Gordon equation. 
The quantum field is expanded in these modes as in (\ref{modeexp}),
and the vacuum is defined as in (\ref{vacdef}).

An important observable is the two point Wightman function
\begin{equation}
G^{W}(X,Y)= \langle{\rm vac} |\phi(X)\phi(Y) |{\rm vac}\rangle~.
\end{equation}
The Wightman function is essentially the response function of 
an Unruh detector, {\it i.e.} a detector carried by a comoving observer;
so it is indeed a physical observable, at least for timelike separated intervals. 
In the free theory, it determines all other correlators, and thus the full theory. 
It is convenient to split the Wightman function into symmetric and 
anti-symmetric parts
\begin{equation}
G^{W}(X,Y) = {1\over 2}\left[G^{(1)}(X,Y) + iD(X,Y)\right]~,
\label{G1Ddef}
\end{equation}
known as the Hadamard function and the commutator function, respectively. 
They have the mode expansions 
\begin{equation}
G^{(1)}(X,Y) = \langle {\rm vac}|\{\phi(X),\phi(Y)\} |{\rm vac}\rangle=
\sum_{n}[u_{n}(X)u^{*}_{n}(Y)+u_{n}^{*}(X)u_{n}(Y)]~, 
\label{G1modes} 
\end{equation}
and
\begin{equation}
iD(X,Y) = \langle {\rm vac} |[\phi(X),\phi(Y)] |{\rm vac}\rangle=
\sum_{n}[u_{n}(X)u^{*}_{n}(Y)-u_{n}^{*}(X)u_{n}(Y)]
\label{Dmodes}~.
\end{equation}
As usual, the Feynman propagator is the time-ordered Green's function
\begin{equation}
iG^{F}(X,Y) = \langle{\rm vac} |T[\phi(X)\phi(Y)]|{\rm vac}\rangle = 
{1\over 2}\left[G^{(1)}(X,Y) + i\sgn(X,Y)D(X,Y)\right]~.
\label{GFdef}
\end{equation}

\subsection{The Euclidean Vacuum}
In the Euclidean vacuum, the modes $\{ u_{n}\}$ are chosen as the {\it regular}
solutions of the Klein-Gordon equation on Euclidean de~Sitter space, {\it i.e.}
the sphere. The correlation functions can then be computed from (\ref{G1modes}-
\ref{Dmodes}), by carrying out the mode sums and then continuing back to
Lorentzian signature.
This procedure gives the generic two-point function\footnote{This is
proportional to the Gegenbauer function $C^{\frac{d-1}{2}}_{-h_+}(-Z).$}
\begin{equation}
f(Z) \equiv C_{d,\nu}F(h_{+},h_{-},{d\over 2},{1+Z\over 2})~,
\label{fdef}
\end{equation}
where $F$ is the hypergeometric function, the weights are
$h_{\pm}={d-1\over 2}\pm i\nu$,
and
\begin{eqnarray}
\nu &=& \sqrt{m^{2} - \left({d-1\over 2}\right)^{2}}~,
\label{nudef} \\
C_{d,\nu} &=& {\Gamma(h_{+})
\Gamma(h_{-})\over (4\pi)^{d/2}\Gamma({d\over 2})}~.
\label{cdnudef}
\end{eqnarray}
The function (\ref{fdef}) is real in the spacelike region $Z<1$, and it has a
pole on the lightcone $Z=1$, which extends to a cut for lightlike $Z>1$ (except
when the $h_{\pm}$ are real and integral.) It is analytic in the complex plane
away from the cut. The normalization constant $C_{d,\nu}$ is set by the
condition that the pole at $Z=1$ has unit strength, as it must be after the
canonical commutation relations have been properly imposed.

The various correlation functions are distinguished by the prescription along 
the cut. The Hadamard function is real and so defined by taking the average 
across the cut
\begin{equation}
G^{(1)}_{0}= f(Z+i\epsilon)+f(Z-i\epsilon) = 2~{\rm Re} f(Z)~.
\label{G1def0}
\end{equation}
The Wightman function has a definite ordering of the fields; so 
its expansion involves the modes $u_{n}(X)$, but the complex conjugate 
modes $u^{*}_{n}(Y)$. The prescription regulating the corresponding infinite 
sum is
\begin{equation}
G^{W}_{0} = f(Z-i\epsilon\sgn(X,Y))~.
\label{GWdef0}
\end{equation}
The Wightman function depends only on the invariant distance 
$Z=-X\cdot Y$, except for the allowance for time-ordering. This
property assures de~Sitter invariance of the vacuum. 
The commutator function is determined from (\ref{G1Ddef}) as
\begin{equation}
D_{0} = 2{\rm Im} ~G^{W}_{0} = 2{\rm Im} ~f(Z-i\epsilon\sgn(X,Y))~.
\label{D0exp}
\end{equation}
This expression vanishes for spacelike separations $Z<1$, as it should
to comply with microcausality.
Comparing (\ref{G1Ddef}) and (\ref{GFdef}), we find the Feynman propagator
\begin{equation}
iG^{F}_{0}= f(Z-i\epsilon)~.
\label{GFdef0}
\end{equation}
The subscript $0$ on each of these correlation functions indicates the 
Euclidean vacuum. 

The index $\nu$ (\ref{nudef}) is real for $m^{2}>({d-1\over 2})^{2}$ and purely
imaginary for $m^{2}<({d-1\over 2})^{2}$. The nature of the corresponding wave
functions differ qualitatively: the ``large mass'' case looks wavy, whereas the
``small mass'' wave functions are exponentially damped. The corresponding
representations of the de~Sitter group are known as the {\it principal} series
and the {\it complementary} series, respectively\footnote{The important special
case $m^{2}=0$ requires different types of representations. The wave functions
given above become trivial in this case; so our discussion is incomplete. For a
recent treatment and original references see \cite{Tolley:2001gg}.}.

The formula (\ref{fdef}) for the function $f(Z)$ simplifies in the 
conformally coupled massless case, equivalent in de~Sitter space to the 
effective mass $m^{2}=d(d-2)/4$. This mass is in the range corresponding to 
the complementary series. The $d=4$ example 
\begin{equation}
f(Z) = {1\over 8\pi^{2}}~{1\over Z-1}~,
\label{confcase}
\end{equation}
just has a simple pole, rather than a cut. The correlation functions
(\ref{G1def0},\ref{GWdef0},\ref{GFdef0}) then simplify correspondingly and
(\ref{D0exp}) gives
\begin{equation}
D_0 (X,Y) = {1\over 4\pi} ~\sgn(X,Y) ~\delta(Z-1)~.
\end{equation}
These expressions are much more transparent to work with than the generic ones,
involving the hypergeometric functions. We will therefore use this special case
in explicit computations.

\subsection{The MA-vacua}
Alternative choices of the modes $\{u_{n}\}$ give rise to different vacua. The
most general option is given by the Bogoliubov transformation 
\begin{equation}
{\tilde u}_{n}(X) = \sum_{n^{\prime}} \left[
A^{*}_{nn^{\prime}}u_{n^{\prime}}(X) - 
B^{*}_{nn^{\prime}}u^{*}_{n^{\prime}}(X)\right]~,
\label{bogtransf}
\end{equation}
of the Euclidean modes. The commutation relations of the QFT
impose the normalization conditions
\begin{equation}
AA^{\dagger} - BB^{\dagger} = I~,
\end{equation}
on the matrices $A$, $B$; and de~Sitter invariance requires each to be 
proportional to the identity matrix $I$. The Bogoliubov transform 
(\ref{bogtransf}) therefore takes the form
\begin{equation}
{\tilde u}_{n}(X) = \cosh\alpha ~u_{n}(X) + \sinh\alpha~ 
e^{i\beta}~u_{n}^*(X)~,
\end{equation}
up to an irrelevant overall phase. With the choice of basis in which
$u_n^*(X)=u_n(X_A),$ the corresponding correlation functions 
can be expressed in terms of the Euclidean ones as
\begin{equation}
G^{(1)}_{\alpha,\beta}(X,Y) = \cosh 2\alpha ~G^{(1)}_{0}(Z) + 
\sinh 2\alpha~ \left[\cos\beta G^{(1)}(-Z) - \sin\beta 
D_{0}({X_A},Y)\right]~,
\end{equation}
and
\begin{equation}
D_{\alpha,\beta} (X,Y) = D_{0}(X,Y)~.
\end{equation}
So the vacua parameterized by $(\alpha,\beta)$ are indeed  invariant under
proper de~Sitter transformations. These are the MA-vacua.

The Euclidean commutator function $D_{0}$ changes sign under time-reversal
$T(X_{0},\vec{X})=(-X_{0},\vec{X})$ so the nonstandard Hadamard 
functions transform as
\begin{equation}
G^{(1)}_{\alpha,\beta}(T(X),T(Y)) =
G^{(1)}_{\alpha,-\beta}(X,Y)~.
\end{equation}
This means the MA-vacua violate CPT invariance for $\beta\neq 0$. 

The Feynman propagators in the MA-vacua are (for Z real)
\begin{equation}
G^{F}_{\alpha,\beta}(Z) = \cosh^{2}\alpha\, G_{0}^{F}(Z)- 
\sinh^{2}\alpha (G_{0}^{F}(Z))^{*} + \sinh 2\alpha
\left({1\over 2}e^{i\beta}G_{0}^{F}(-Z) + {\rm c.c.}\right)
\label{maprop}
\end{equation}
and similarly for the Wightman function $G^{W}_{\alpha,\beta}(Z)$.
In the conformally invariant case (\ref{confcase})
\begin{equation}
iG^{F}_{\alpha,0}(X,Y) = {1\over 8\pi^{2}}[\cosh^2\alpha {1\over Z-1-i\epsilon}
+\sinh^2\alpha {1\over Z-1+i\epsilon} + \sinh 2\alpha{1\over Z+1} ]~,
\label{mapropc}
\end{equation}
for $\beta =0$. This expression has several noteworthy features
\begin{itemize}
\item
The strength of the pole along the lightcone $Z=1$, and thus the
ultraviolet divergences, are larger than they would be in 
flat space. The renormalization program can therefore not rely on the
flat space limit to determine the appropriate counterterms.
\item
There is a pole along the lightcone $Z=-1$ emanating from the antipodal 
event. Such correlations will appear as acausal effects in some 
processes (see section \ref{sect:treeampl}) . 
\item
The two terms having poles along the light-cone $Z=1$ have {\it opposite}
$i\epsilon$ prescriptions. Similarly, the pole along the lightcone of the 
antipodal event $Z=-1$ is understood as a principal value, {\it i.e.}
the average of the two $i\epsilon$ prescriptions with opposite sign.
This structure will render loop amplitudes ill-defined (see
section \ref{sect:loopampl}).
\end{itemize}
These features are not merely unfamiliar, they prevent a definition of the 
theory at the interacting level. 

The MA-vacua can formally be represented as squeezed states in the 
Euclidean vacuum
\begin{equation}
|\alpha,\beta\rangle = {\cal N}~\exp \left( -{1\over 2}\sum_{n}[\alpha 
e^{-i\beta}(a_{n}^{\dagger})^{2} - {\rm c.c.}] \right) | 0 , 0 \rangle~.
\end{equation}
According to this representation the MA-vacua can be interpreted
as minimal uncertainty states with the field in opposite ends of the 
universe correlated precisely at the quantum level.
Of course, these squeezed states are not normalizable and therefore 
the $|\alpha,\beta\rangle$ are actually not elements in the 
Hilbert space of the Euclidean vacuum $|0,0\rangle$. 
That is one reason why the MA-vacua really are new vacua, and not just a special class of states. 

\section{Interacting Quantum Fields}
\label{sect:interacting}
In this section we first discuss the general issues of observables in de~Sitter
space, Feynman rules in curved spacetimes, and the singularities of QFT
amplitudes. We then apply these remarks to the MA-vacua, discussing in turn the
tree amplitudes and loops. The section ends with comments on axiomatic QFT in
the Euclidean vacuum.

\subsection{Fundamentals}

One question that arises in curved spacetimes is what can be 
observed.  In general, there are no asymptotically-flat regions in 
which to define in- and out-states.  Thus, there is no well-defined 
S-matrix, although in de~Sitter space, there have been discussions of 
meta-amplitudes\cite{Witten,SV} for transitions between states 
defined on the past null infinity ($\ical^-$) and the future null 
infinity ($\ical^+$).  For our purposes, it is sufficient to imagine 
observers carrying calibrated detectors, idealized as Unruh 
detectors, that can be used to probe interactions with the field on 
their (timelike) trajectories.  The 
corresponding ``detector response function" probes the two-point 
correlation function (or Wightman function) of the  
field. (This has been reviewed in refs~\cite{Birrell:ix,SSV,BMS}.) 
Going beyond the usual linear response analysis, it should be possible 
to measure higher point correlators as well. Moreover, further 
correlations can be accessed by introducing several detectors 
or independent observers. We will assume that {\it all Wightman 
functions are in principle observable, at least for points separated by 
timelike geodesics}.

In some applications, the arguments of the correlation functions should 
be restricted to take values within a given static patch. For 
example, the complementarity principle can be taken to assert that, in the 
complete quantum theory of gravity, the full set of correlators, 
restricted in this way, form a complete set of observables\cite{complcits}. 
This is consistent with our interpretation of correlators, motivated by 
effective field theory. Indeed, ``complementarity'' precisely 
states that the fundamental, holographic view of observables is consistent 
with the low energy, local interpretation of physics. In the present 
context the latter is more appropriate.

The next question is how to get the Feynman rules for calculating 
correlation functions perturbatively.  In fact, moving beyond free 
field theory to interacting field theories in a fixed, curved 
background is fairly straightforward.  Formally, we may define the 
Feynman rules for perturbation theory in the same way that it is done 
in Minkowski space via the Feynman path integral (FPI).
\beq\label{fpi}
 e^{iW[J]}=e^{i\int dx\sqrt{g} \lcal_I(-i\frac{\delta}{\delta 
J})}e^{iW_f[J]}~,
\eeq
where $W[J]$ is the generating functional of connected Green's functions, and
$\lcal_I(\phi)$ is the interaction Lagrangian density (assumed to be
nonderivative).  $W_f[J]$ is the free field generating functional given by
\beq\label{freeW}
W_f[J]=\half\int dx \sqrt{g(x)} dy \sqrt{g(y)}J(x) G^F_\ab(x,y) J(y)~.
\eeq
In Minkowski space, the FPI underlying this expression is normally {\it defined}
by its Euclidean counterpart\cite{coleman}. Since, in the MA-vacua, the
propagators are not analytic and do not permit a Wick rotation, we cannot employ
such a convention here but instead wish to work directly in Lorentzian
signature.  A common prescription is simply to replace $m^2$ by $m^2- i\epsilon$
to achieve convergence of the FPI.  Alternatively, the expressions (\ref{fpi})
and (\ref{freeW}) make sense formally without any reference to functional
integration. Thus, the coordinate space Feynman rules in general curved
spacetime have the same general form as in flat space but with a modified
propagator and with insertion of the appropriate factors of the background
metric.  These observations apply to any quantum field theory (QFT) in curved
spacetime, not just to de~Sitter space.

According to these rules Feynman amplitudes are written as integrals over
various products of propagators as determined by the vertices. We are interested
in the nature of singularities of such expressions.  Propagators are
distributions rather than functions, \ie, they become singular for certain
values of their arguments. Generally, $n$-point correlation functions, in the
physical region, are also distributions rather than real functions.
Distributions make sense as linear functionals when integrated over certain test
functions.  However, it is by no means obvious that Feynman amplitudes,
involving products of propagators and consequently products of distributions,
are well-defined.  

For the usual vacuum in Minkowski space, the situation is well-understood
\cite{ELOP}. All Wightman functions are boundary value distributions of analytic
functions.  All Feynman integrals can, by a Wick rotation to Euclidean
signature, be written in such a way that the integrands have no singularities at
all. Renormalization may be carried out for Euclidean signature, so the
renormalized $n$-point functions are well-defined and nonsingular for Euclidean
signature.   Upon analytic continuation back to Minkowski space, singularities
may be encountered, but only when certain ``causal, on-shell" conditions are
satisfied.  The upshot is that all  renormalized $n$-point functions are well-
defined functions in the physical region for Lorentzian signature, except  for
singularities that correspond to physically allowed processes.  

It seems that this program may also be extended to de~Sitter space in the
Euclidean vacuum\cite{Bros:1990cu,Bros:dn,Bros:1995js,Bros:1998ik}. However, in the 
MA-vacua, the propagators are very different distributions whose singularities
prevent analytic continuation. As a result, it is not at all clear that a
sensible perturbation expansion can be associated with the formal Feynman rules.
One must address old questions anew, such as, when do the singularities of
integrands of Feynman amplitudes correspond to singularities of the integrals?  

\subsection{Tree Amplitudes}
\label{sect:treeampl}
We first show that, at tree level, all Feynman amplitudes are in fact
mathematically well-defined. Indeed, consider a particular perturbative
contribution to an $n$-point function $G_\ab^{(n)}(x_1,x_2,\ldots, x_n)$ associated
with the time-ordered product of fields.  Propagator singularities correspond to
hypersurfaces in the multidimensional spacetime of the integrand of the Feynman
amplitude.  For {\it any}\/ Feynman diagram, one may choose values for the external points so that the external propagators are nonsingular at points of integration where internal propagators diverge.  So one need not worry about coincident singularities involving external legs; it is only the truncated $n$-point functions that could cause problems.\footnote{By ``trunctated," we mean with external propagators removed by multiplying by the inverse propagator, \ie, applying the wave equation to the external legs.} Tree diagrams are special inasmuch as the truncated amplitude involves no internal integrations and, for generic external points, there are no propagators having coincident singularities (such as squares of propagators as will appear in loops.)  Therefore, the truncated diagram is well-defined except for special values of the external points.  When one reattaches external legs, the new integrations will not produce coincident singularites except possibly for special values of the external points.  One need only show that the full Green's function is independent of the order in which the external legs are reattached, which is the case.  In other words, the definition of propagators as linear functionals suffices to define tree diagrams so, in tree approximation, the $n$-point function is well-defined for some range of external points.  

On the other hand, an amplitude thus defined clearly has singularities 
for particular external
points and, in order to have a properly defined theory, those singularities must
have a sensible physical interpretation. In the MA-vacua, the singularities in
the propagators associated with the antipodal point give rise to seemingly
unphysical singularities in amplitudes which are a serious cause of concern. A
possible resolution of this problem is that the antipodal points are somehow
``hidden'' behind a horizon but, as we show next, in the interacting theory
these singularities cannot be so easily hidden.    
\begin{figure}
\center{
\epsfig{file=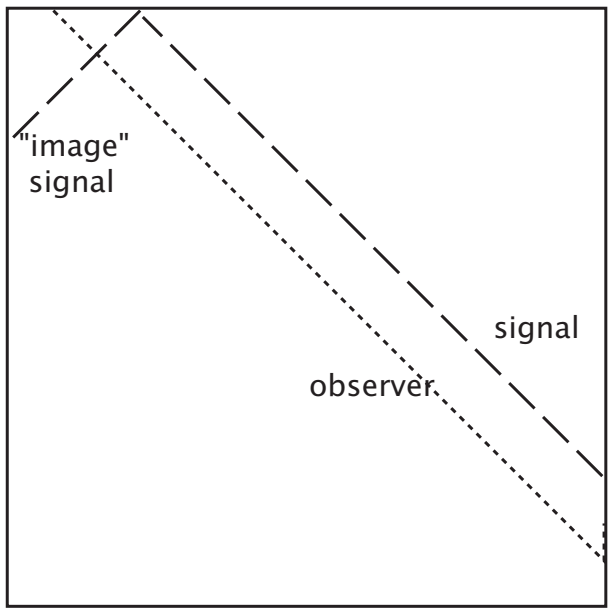, height=8cm, width=8cm}
}
\caption{Causal relations to the antipodal point.}
\label{fig3}
\end{figure}
Consider generating a signal at the south pole, propagating towards the north
pole, and a corresponding ``image'' disturbance from the antipodal point
propagating from the north pole (see fig.~\ref{fig3}.)\footnote{Note that the
``image'' signal satisfies the homogeneous Klein-Gordon equation so it is not
generated by a physical source at the north pole.} The respective lightcones
meet only in the asymptotic future, at ${\cal I}^{+},$ but this does not mean the
``image'' is unobservable. Any observer sent off (at the speed of light) {\it
prior} to generating the signal {\it will} experience the ``image''.
Alternatively, when gravitational back-reaction is taken into account, the
causal diagram of de~Sitter space will become slightly
``taller''\cite{Leblond:2002ns}, and so the signal and its ``image'' can meet.
These simple examples suggest that the antipodal singularities are in fact a
cause for concern in the interacting theory.
\break

\begin{figure}
\center{
\epsfig{file=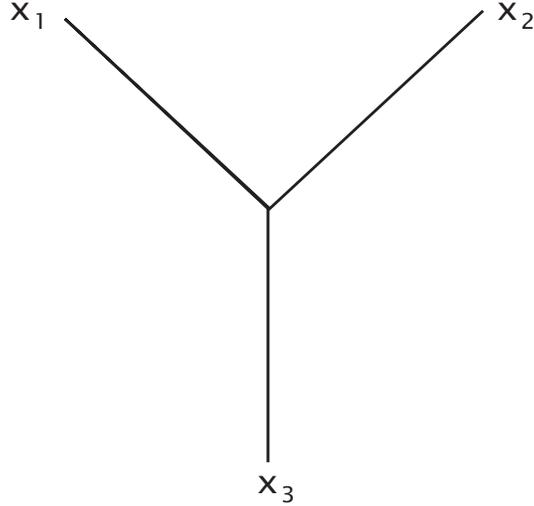, height=7cm, width=7cm}
}
\caption{Simple tree diagram.}
\label{fig1}
\end{figure}
\begin{figure}
\center{
\epsfig{file=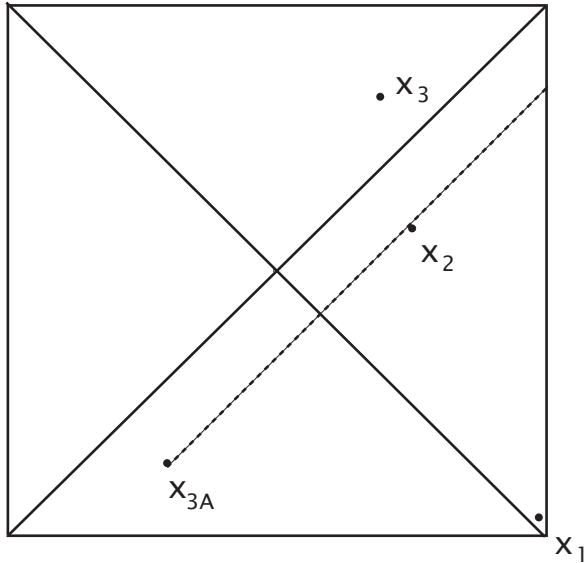}
}
\caption{Penrose diagram for three points.}
\label{fig2}
\end{figure}

\nobreak
To verify this expectation in perturbation theory, consider the simple case of a
$\lambda\phi^3$ interaction and the three-point function (fig.~\ref{fig1})
with amplitude
\beq\label{threepoint}
G^{(3)}(x_1,x_2,x_3)=-\lambda\int dy \sqrt{g(y)} G^F_\ab(x_1,y) 
G^F_\ab(x_2,y) G^F_\ab(x_3,y)~.
\eeq
The integration runs over all of de~Sitter space. Each of the propagator 
factors $G^F_\ab(x_i,y)$  is singular when $y$ 
is on the lightcone of the corresponding point $x_i$ or of its 
antipode $x_{iA}.$ 
Consider the external points $x_i$ situated as in 
the Penrose diagram in fig.~\ref{fig2} 
The event $x_1$ is on 
(or near) $\ical^-,$ and the point $x_3$ is within the causal 
future of $x_1,$ but near the lightcone and just outside the causal 
diamond.  Thus, its antipode, $x_{3A}$ lies just outside the causal 
future of $x_1,$ so that $Z(x_{3A},x_1)<-1.$  Nevertheless, the 
point of integration $y$ may lie on the future lightcone of $x_{3A}$ 
while within the causal diamond of $x_1.$  Moreover, singularities of 
the propagators coincide if $x_2$ also lies on the lightcone of $x_{3A}$.  
In this case the integral, and thus the amplitude, will be singular.
Since $x_{1}$, $x_{2}$, $x_{3}$ can be placed on a nowhere spacelike 
trajectory, we find it difficult to believe that this (singular) correlation would
not be not observable. Moreover, being a tree diagram, the effect 
discussed here is essentially a property of the classical theory. We also note 
these are not short-distance singularities but are associated with all points 
on the lightcones, no matter how distant from the apex.  

These considerations do not prove that the singularities of tree-amplitudes,
however unintuitive and nonlocal,  are absolutely impermissible; but such
singularities certainly would lead to strange events and traumatic experiences.
Their existence raise serious questions about the interpretation, and possibly
even the  viability, of the MA-vacua.

That tree diagrams can be defined is helpful also in understanding free field 
theory.  One could of course regard the mass term as a two-point interaction 
vertex and build up the full propagator by summing up this interaction.  For
consistency, one ought to get the same answer as in (\ref{maprop}).  The 
fact that tree amplitudes are well-defined ensures that this can be done, and 
the result will be unambiguous.  Unfortunately, this conclusion does not 
extend to loop amplitudes, to which we now turn.

\subsection{Loop Amplitudes}
\label{sect:loopampl}
The analysis of higher order corrections, involving loop amplitudes, 
is more complicated but also more bedamning.  When the 
propagators are boundary values of 
analytic functions, as in the Euclidean vacuum in Minkowski or 
de~Sitter space, the general conditions under which a singularity of 
the integrand actually results in a singularity of the integral have 
been thoroughly analyzed and are reviewed, for example, in 
\cite{ELOP}.  In Minkowski space, the singularities are 
generally discussed in momentum space, where they have been codified 
by the Landau rules\cite{Landau}, or by a related prescription, 
Cutkosky's ``cutting rules"\cite{Cutkosky} for putting 
internal particles on-mass-shell.  Although this methodology is not 
available for theories in curved spacetime, the same sort of analysis 
{\it can} be applied to a coordinate space formulation.   We shall 
illustrate this in the case of the self-energy diagram below, but
first we shall try to provide an overview of the issues.  

In general, a singularity of the integrand is not a singularity of the integral
if the integrand is sufficiently analytic in a neighborhood of the singularity
so that the path of integration can be deformed and moved away from the
singularity.  In the Euclidean vacuum 
where the propagators are boundary values of analytic functions,
singularities of the integrand do not generally yield singularities of the
integral because, loosely speaking, the propagators involve a consistent
prescription with all masses having infinitesimal negative imaginary parts (the
familiar $-i\epsilon$ prescription).  The most common exception that produces a
singularity of the integral, and the one of particular interest in the present
context, is when the contour is ``pinched" between singularities of the
integrand that coalesce from opposite sides of the integration contour,
preventing its deformation away from the singularities.  Normally, the pinch
only occurs when more than one propagator are simultaneously singular. Even in
the Euclidean vacuum, since propagators are distributions rather than functions,
it is by no means obvious that products of singularities make sense.  Once
again, analyticity comes to the rescue, because for Euclidean signature, the
integrands are nonsingular, and, once again, it is possible to show that the
integrals in the physical region are boundary values of analytic functions.
However, for the MA-vacua, where the propagators do not have such analyticity,
this pinching occurs for each individual propagator.  When singularities of
different propagators coincide, it seems to be doubtful that a well-defined
determination of the Feynman integral can be made.

\begin{figure}
\center{
\epsfig{file=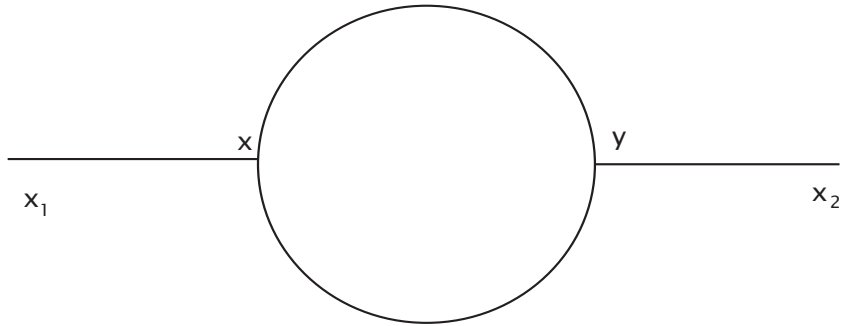, height=4.25cm, width=11.5cm}
}
\caption{Self-energy diagram.}
\label{fig4}
\end{figure}

We shall show next that even the simplest of loop diagrams is not 
well-defined for the MA-vacua.  The self-energy diagram of 
fig.~\ref{fig4} is one of the most elementary loop diagrams occuring in 
perturbation theory.  
The associated expression is 
\beq\label{selfenergy}
-\lambda\int dx\sqrt{g(x)} dy\sqrt{g(y)} G_\ab^F(x_1,x) 
G_\ab^F(x,y)^2 G_\ab^F(y,x_2)~,
\eeq
involving the square of the internal propagator. 
Although the self-energy
requires renormalization, we shall assume that appropriate counterterms have
been included.\footnote{It should be noted that, in the MA-vacua, the
relationship between real and imaginary parts of the propagators is nonstandard.
We have not analyzed the ultimate effects this might have on the renormalization
program.  That too may cause problems, but the point we wish to make is not
exclusively a short distance issue, so be reconciled by some cutoff or softening
of the theory.  $G^F_\ab(x,y)$ is singular whenever $y$ is on the lightcone 
of either $x$ or $x_A.$} 
\nobreak
\nobreak
In Minkowski space, the loop does not
contribute a singularity to the self-energy unless the pair $(x,y)$ are such
that the particles in the loop are real, that is, on-mass-shell, propagating
forward in time\cite{C-N}.  Otherwise, because of the analyticity of the
Feynman propagators, the potential singularities of the integrand can be
avoided.  Another way to state the same result is that the Green's functions are
all nonsingular for Euclidean signature, and the only singularities encountered
in analytic continuation to the physical region arise for
classically realizable processes.  In de~Sitter space, the propagator is not
analytic in the MA-vacua but only in the Euclidean vacuum.
Moreover, the propagator is a distribution whose square is ill-defined, \ie,
the square of principal values and delta-functions are not defined.  We simply
do not know how to define this integral\footnote{One might imagine that there
is some sort of cancellation between the principle values at $Z=1$ and $Z=-1$,
but one may verify that this does not happen.}. It seems incumbent upon those
who propose to work in these non-standard vacua to explain the rules of
calculation and their interpretation. 

Although our arguments are generic, we shall illustrate the problems in the
specific case of the massless, conformally coupled scalar\footnote{For purposes
of this discussion, we assume also that a finite mass counterterm has been added
so that the renormalized mass corresponds to the conformally massless case.},
for which the various two-point functions  reduce to poles at $Z=\pm1$, their
principal values, or imaginary parts.  For that case, the discussion is not
obscured by the complications of dealing with hypergeometric functions.

\begin{figure}
\center{
\epsfig{file=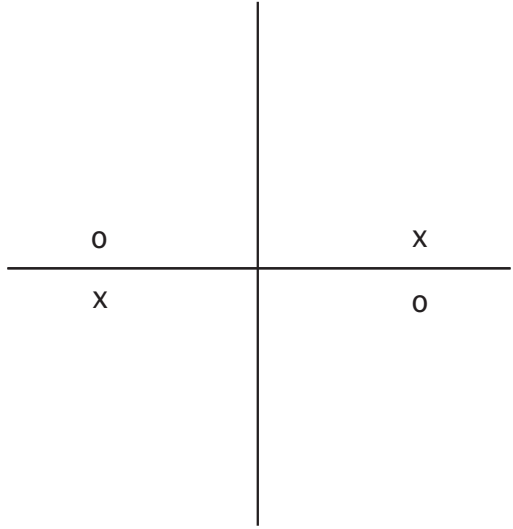, height=7cm, width=8cm}
}
\caption{Pinched singularities.}
\label{fig5}
\end{figure}

Before discussing the de~Sitter case, let us first remind ourselves of the
situation in flat space.  The one-particle-irreducible (1PI) self-energy 
\beq\label{sigma}
\Sigma(x,y)\propto \left[\frac{1}{(x-y)^2-i\epsilon}\right]^2.
\eeq
This highly singular expression is manageable inside integrals such as
(\ref{selfenergy}), because its singularities lie on the same side of the
integration contour. If we imagine performing the integration over $x^0$ or
$y^0$, then we have $x^0-y^0=\pm(|\vec{x}-\vec{y}|+i\epsilon$), 
as depicted by the crosses in fig.~\ref{fig5}.  
This does not produce a singularity as $\epsilon\to0$ unless it occurs
in an integral in which another singularity prevents deforming  the path of
integration.  The full expression, (\ref{selfenergy}), involves\footnote{Of
course, in flat space, $G^F$ is a function of the difference $x_1-x_2$ only.}
\beq\label{selfenergytwo}
G^F(x_1,x_2) \propto \int dx dy \frac{1}{(x_1-x)^2-i\epsilon} 
\Sigma(x,y) \frac{1}{(y-x_2)^2-i\epsilon}.
\eeq
Since the propagators for the external legs depend on the external points, their
singularities will not in general coincide with those of $ \Sigma(x,y).$  Thus,
it is only  special values of $(x_1-x_2)$ that produce a singular integral.  In
this case, by the Coleman-Norton theorem\cite{C-N}, these occur only when
$(x_1-x_2)$ is a null vector, and $x$ and $y$ are proportional to $(x_1-x_2),$
so that $(x_1-x)$ and $(y-x_2)$ are also null\footnote{For the 
massless case, one may show this directly in coordinate space by introducing
Feynman parameters as is usually done in momentum space.}.  

Now consider the corresponding situation in de~Sitter space in an MA 
vacuum, with 
\beq
\Sigma(x,y)\propto \left(G_\ab^F\right)^2,
\eeq
where the propagator $G_\ab^F$ is given in (\ref{maprop}).  This square involves
a great many terms and is rather complicated.  We shall simplify the discussion
in two inessential ways:  we restrict our attention to the CP-invariant vacua,
$\beta=0,$ and we focus on the conformally massless case (\ref{mapropc}).  Then,
the 1PI-self-energy involves terms of various sorts.  From the terms with
singularities at $Z=1,$ we get a cross-term of the form
\beq\label{crossterm}
\frac{\sinh^\alpha 
\cosh^2\alpha}{(Z(x,y)-1-i\epsilon)(Z(x,y)-1+i\epsilon)},
\eeq
where 
\beq
Z(x,y)-1=\half (X-Y)^2 = 
\frac{(\eta_x-\eta_y)^2-(\vec{x}-\vec{y})^2}{2\eta_x\eta_y}.
\eeq
The first expression involves the embedding coordinates associated 
with the points $x$ and $y,$ while, in the second form, we have 
represented this factor in planar coordinates\cite{SSV}, since the 
denominators then look the same as in flat space.  (In planar 
coordinates, the metric is conformally flat, taking the form 
$$ds^2=\frac{1}{\eta^2} \left(d\eta^2-d\vec{x}^2\right).$$
Although the coordinates are singular at $\eta=0,$ this is of no consequence
for the discussion of the singularities of the self-energy.)  Because the
denominators in (\ref{crossterm}) involve opposite signs of $i\epsilon,$ there
are now complementary singularities denoted by the circles in fig.~\ref{fig5}, so
the singularity at $Z=1$ pinches the contour regardless of the values of the
external points.  Even if such a peculiar singularity were somehow physically
acceptable, what is the value of an integral in which it appears?  To be the
principal value, there would have to be a factor of $Z-1$ in the numerator; to
be a delta-function, there would have to be a factor of $\epsilon$ in the
numerator.  As it is, it lacks definition, and the self-energy is ill-defined.
This cannot be cured by local counterterms in the Lagrangian; it happens
everywhere along the lightcone.  A similar thing is true of the
contribution of the antipodal singularity at $Z=-1.$  It is initially defined as
the principal value in (\ref{mapropc}), so its square is ill-defined for the
same reason.  The CPT-noninvariant case, $\beta\ne0,$ introduces further
conundrums.  

The only case that avoids such problems is the Euclidean vacuum, 
$\alpha=\beta=0,$ for which there is a consistent sign to the 
$i\epsilon$ assignment, just as in Minkowski space.  The discussion 
of singularities carries over to the de~Sitter case, 
complicated only by having just a coordinate space rather than a 
simple momentum space representation of the propagators. In the 
general case other than  the massless, conformally coupled scalar, 
the discussion is more complicated because, in addition to these pole 
terms, there are also branch points to be dealt with.  But the 
outcome is the same.\footnote{In odd dimensions, the coordinate representation
of propagators has a ``kinematical" branch point associated with certain
square roots that need to be treated rather differently from dynamical
singularities.}  
In curved spacetime, it seems that a necessary condition for an interacting QFT,
at least one that has a well-defined perturbation expansion, is that the $n$-
point functions be boundary values of analytic functions.  

\subsection{Axiomatic QFT}
\label{sect:axiomatic}
There are additional reasons for choosing the Euclidean vacuum for de~Sitter
space.  J. Bros and collaborators,\cite{Bros:1990cu,Bros:dn,Bros:1995js,Bros:1998ik} 
taking an axiomatic approach, have shown that most of the
properties of QFT familiar from Minkowski space can be carried over to
de~Sitter space.  This requires one crucial change in the usual axioms.
Normally, analyticity of Wightman functions is derived from the assumption of a
Hamiltonian with a positive spectrum.  In the de~Sitter case, there is no
Hamiltonian, so Bros \etal\ simply assume that the correlations functions in
coordinate space may be extended to complex de~Sitter spacetime, so that they
may be associated with distributions that are boundary values of analytic
functions\footnote{This will be true for any QFT in curved spacetime in which
the Wick rotation from Euclidean to Lorentzian signature can be carried out.}.
With this assumption, Bros \etal\ derive a K\"all\'en-Lehman representation
with positive spectral function \cite{Bros:1995js}, extend the Bisognano-
Wichmann theorem\cite{BW}, prove the KMS property required for a thermal
interpretation of the two-point function \cite{Bros:1998ik}, as well as the CPT
theorem  and the Reeh-Schleider theorem\cite{RS,SW}. This last property is
especially important since, classically, no single observer can see all of
de~Sitter space.  Loosely, this theorem states that knowledge of the Wightman
functions on an open subspace of spacetime implies knowledge of the functions
everywhere.  In other words, the field theory is uniquely defined by its action
within a restricted domain, for example, within the causal diamond of a single
observer.  The Reeh-Schleider theorem then assures us that the QFT is uniquely
defined everywhere.  Thus, analyticity is an extremely powerful tool, whose
potential for the analysis of field theory in curved spacetime deserves greater
study.  Even for generalized free fields, the axioms are satisfied only by the
Euclidean vacuum.

\section{Thermal Properties}
\label{sect:thermal}
The MA vacua are not thermal. The purpose of this section
is to characterize their state more precisely and discuss the
role of interactions from this point of view.

\subsection{The Free Theory}
A comoving observer in de~Sitter space follows a path parameterized
as $X_{d}=\cosh\tau$, $X_{0}=\sinh\tau$ where $\tau$ is the proper
time. The invariant distance between the observer and a reference
event at $\tau=0$ then evolves as $Z=\cosh\tau$. The transition
rates in a detector following such a trajectory are proportional to 
the response function
\begin{equation}
F_\ab(\Delta E) = -\int_{-\infty}^{\infty} d\tau e^{-i\Delta E\tau}G_\ab^{W}(\tau)~,
\end{equation}
where $\Delta E= E_{j}-E_{i}$ is the energy difference between any two levels
of the detector. 

In the simple example of the conformally coupled field in four 
dimensions, the Wightman function becomes
\begin{equation}
G^{W}_{\alpha,\beta}(\tau)= 
{1\over 4\pi^{2}}~
\left( {\cosh^{2}\alpha\over (2\sinh {\tau-i\epsilon\over 2})^{2}}+
{\sinh^{2}\alpha\over (2\sinh {\tau+i\epsilon\over 2})^{2}}
+ {\sinh 2\alpha \cos\beta\over 2\cosh^{2}{\tau\over 2}}
\right)~,
\label{cfw}
\end{equation}
and the corresponding response function can be computed using contour 
integration. The result is
\begin{equation}
F_{\alpha,\beta}(E) = {E\over 2\pi}~{1\over e^{2\pi E}-1}~|\cosh\alpha
+ \sinh\alpha e^{i\beta} e^{\pi E}|^{2}~.
\label{respfct}
\end{equation}
The dependence of the response function on $(\alpha,\beta)$ 
\begin{equation}
F_{\alpha,\beta}(E) = F_{0}(E)~|\cosh\alpha
+ \sinh\alpha e^{i\beta} e^{\pi E}|^{2}~,
\end{equation}
in fact follows from the analytic structure of $G^W_\ab$ in proper time
and is valid for arbitrary dimension and general mass.  
This can be shown from the KMS condition (discussed in more detail below) 
\begin{equation}
G^{W}_{0}(-\tau-2\pi i)= G^{W}_{0}(\tau)~,
\label{KMS}
\end{equation}
satisfied by the Euclidean vacuum, as well as the reality condition
\begin{equation}
G^{W}_{0}(\tau)^{*} = G^{W}_{0}(-\tau)~,
\end{equation} 
for real $\tau$, by generalizing the derivation given, \eg, in \cite{BMS}. 

Is this response function compatible with thermodynamic equilibrium? Denoting
the occupation number of level $i$ by $N_{i}$, and the transition probability
between levels $i$ and $j$ by $P_{ij}$, the condition for equilibrium in the
detector is
\begin{equation}
\sum_{i\ne j} \left[ N_{i} P_{ij} - N_{j} P_{ji} \right] = 0~,
\label{eqcond}
\end{equation}
for each level $j$. One way to obtain equilibrium is if the expression in square
brackets vanishes for each pairs $i,j$; this is the {\it detailed balance} condition.
It implies that the ratio of probabilities 
\begin{equation}
{P_{ij}\over P_{ji}} = {N_{j}\over N_{i}}~,
\label{factor}
\end{equation}
factorize into expressions depending only on the states $i$ and $j$
but not on both. In the present context
\begin{equation}
{P_{ij}\over P_{ji}} = {F(\Delta E)\over F(-\Delta E)}  = 
e^{-2\pi\Delta E}
\left| {\cosh\alpha + \sinh\alpha e^{i\beta} e^{\pi \Delta E}\over
\cosh\alpha+ \sinh\alpha e^{i\beta} e^{-\pi \Delta E}}\right|^{2}~.
\label{ratprob}
\end{equation}
For $\alpha=0$ this expression factorizes as (\ref{factor}) with $N_{i}\propto 
e^{-2\pi E_{i}}$ but otherwise it does not factorize at all. Thus the principle of
detailed balance is violated for $\alpha\neq 0$.\cite{SSV,BMS} 

The equilibrium condition (\ref{eqcond}) is more complicated
when detailed balance is violated. Defining 
\begin{equation}
P_{jj}\equiv -\sum_{i\ne j}P_{ji}~,
\label{piidef}
\end{equation}
so that (\ref{eqcond}) may simply be written as 
\begin{equation}
\sum_{i}  N_{i} P_{ij}=0, 
\end{equation}
where the sum now extends over all $i$.  Then there is a solution for the
$N_{i}$ if and only if the determinant of the matrix with elements $P_{ij}$
vanishes.  It is easy to see that, in fact, this is always true, since the
columns of $P_{ij}$ are not linearly independent due to the definition
(\ref{piidef}). Thus, {\it a detector in the MA-vacua equilibrates, but it does
not satisfy detailed balance}.

The ratio of probabilities (\ref{ratprob}) is larger in the MA-vacua than in the
Euclidean one. This indicates that high energy states are more populated than
they would be in the thermal state. The precise equilibrium distributions in the
MA-vacua are not universal; they depend on properties of the detector. The
reason is that the occupation number of a given level $i$ depends not only on
the energy of that level, but also on the energy of all the other levels. Thus,
if one wants to find the MA equilibrium distribution functions, one must make
assumptions about the available spectrum.

Our analysis is consistent with standard text book discussions of
detailed balance, although the terminology can be confusing. The ``principle 
of microscopic reversability''\footnote{Also called ``detailed
balance" by some authors.} 
of the S-matrix $S_{if}= S_{f^{*}i^{*}}$ follows 
from time reversal invariance and other general principles. A variant of this 
statement is presumably valid in the present context insofar as time reversal is a 
symmetry, {\it i.e.} one must assume $\beta=0$. The principle of detailed 
balance follows from the microscopic irreversability under the additional 
assumption of equipartition, {\it i.e.} all microscopic states are 
occupied with equal probability. The MA-vacua violate detailed balance 
because they correspond to a different distribution, with relatively more weight 
at higher energy. 

\subsection{Interactions}
The discussion of thermal properties has so far ignored interactions. Of course,
thermodynamics crucially relies on interactions for ergodicity and
thermalization of the physical system. This opens the possibility that the 
MA-vacua would in fact equilibrate in the presence of interactions, eventually
approaching the Euclidean vacuum\cite{shenker}. On the other hand, since the
MA-vacua are in fact de~Sitter invariant, with correlations between far flung
regions, it is not clear that interactions, operative only locally, would be
able to thermalize the state. 

In our view, the role of interactions is less
dynamical in the present context and is rather one of principle. As we already
argued in the previous section, it is not clear that it is even possible to
include interactions. If we view an MA-vacua as a thermodynamic system in a
nonstandard equilibrium we can try to include interactions perturbatively,
averaging appropriately over the equilibrium configurations of the system.
Techniques for carrying out this type of computation have been systematically
developed in thermal quantum field theory (TQFT). One of the basic axioms of
TQFT, ensuring the consistency of the theory, is the KMS condition
(\ref{KMS}). We will show below that the KMS condition is satisfied in the 
Euclidean vacuum\cite{Bros:1998ik} 
but it is violated in the MA-vacua. Thus {\it interactions 
can be included systematically in the Euclidean vacuum but, in the MA-vacua, 
it is not known how to compute corrections due to interactions, even in 
principle.} This is how the problems discussed in more detail in the 
previous section reappear from a thermal point of view.

To verify the KMS condition in the Euclidean vacuum, consider for simplicity
the principal series in $d=3$. In this case the hypergeometric function 
simplifies, and the Wightman function in the Euclidean vacuum becomes
\begin{equation}
G^{W}_{0}(\tau)= {i\over 8\pi\sinh \pi\nu}~{e^{-\pi\nu + i\nu\tau}
-e^{\pi\nu - i\nu\tau}\over \sinh\tau}~,
\end{equation}
and (\ref{KMS}) is easily verified. In the MA-vacua, one of the terms
in the Wightman function is the complex conjugate of this expression.
This term is invariant under $\tau\to -\tau + 2\pi i$, but {\it not} under 
$\tau\to -\tau - 2\pi i$, as the KMS condition prescribes. The
full correlation function therefore violates the KMS condition.

It is not difficult consider the full hypergeometric function and thus
extend this argument to masses in the complementary series, and to 
general dimensions. The $d=4$ conformal case (\ref{cfw}) illustrates
the special care needed due to the $i\epsilon$ prescriptions. The 
correlation function is a distribution, {\it i.e.} a linear functional
\begin{equation}
\psi \to \int_{-\infty}^{\infty} d\tau G^{W}(\tau)\psi(\tau)~,
\end{equation}
on a suitable class of test functions $\psi$. The KMS
condition amounts to invariance under a move in the path of integration 
downwards by $2\pi i$, for all $\tau$. 
The first term in (\ref{cfw}) has a pole just {\it above} the real axis, 
so the move is unobstructed. This ensures the KMS condition for the 
Euclidean vacuum. However, the second term in (\ref{cfw}) has a 
pole just {\it below} the real axis, obstructing such a move. This
shows the MA-vacua violate the KMS condition.

Before concluding the paper let us make some remarks on standard TQFT.
Historically TQFT was plagued by pinched singularities and divergences arising
from formal manipulations with distributions, such as squaring $\delta$-
functions. These difficulties were overcome with the advent of the real-time
formalism for thermal field theory (for review see {\it e.g.} \cite{lebellac})
and, given the similarities with the problems we have encountered in de~Sitter
space, one might try to repeat this success here\footnote{We thank J.~Cline and
R.~Holman for this suggestion.}. Time-dependent correlation functions in TQFT
involve questions defined on the real time axis but also, since the theory is
thermal, correlation functions must be periodic in imaginary time. These dual
requirements are accomplished by defining correlation functions on a contour
along the entire real axis, with a return ending up with the same real part as
the starting point, but shifted in the imaginary part as required by
temperature. Since the return path necessarily goes in the ``wrong'' direction
of time this type of correlation function is sensitive to certain ghost-type
fields which, as it happens, cancel the divergences present in more na\"{\i}ve
approaches. The key principle in TQFT is thus periodicity in the imaginary time,
or the KMS condition; but {\it it is precisely the KMS condition that is
violated in the MA-vacua}, leading to difficulties. One could try to treat
$\alpha\over 2\pi$ as a formal temperature but the corresponding periodicity
does not seem related to the imaginary part of a physical time. In our view a
more appropriate use of TQFT would be to the Euclidean de~Sitter vacua, or to
black hole backgrounds. In either case the background is precisely thermal and
it is suggestive that, to study time-dependent correlation functions, we must
take into account a region with ``wrong'' time direction, as well as two
``vertical'' regions, defined with imaginary time. These parts of the
integration path are related to regions of the causal diagram in
spacetime\cite{kos}, and this gives hope that one can understand the role played
by nontrivial causal structure even for processes apparently taking place in a
single causal patch.

\section{Discussion}
\label{sect:conclusion}

We have argued that, of the alternate de~Sitter invariant vacua of Mottola and
Allen, only the Euclidean vacuum has sufficient analyticity to admit a 
well-defined perturbation theory.  The singularities of the propagators in other
vacua not only have ``unphysical" singularities associated with the lightcone
of antipodal points, but also appear not to permit a definition of loop diagrams
in general.  

Analyticity is the main ingredient in our considerations. More generally, we
note that analyticity is a common denominator of sensible field theories in both
flat and curved spacetime: in discussions of the adiabatic vacuum or of
cosmologies having asymptotically flat regions, the preferred vacua lead to
analytic correlation functions and S-matrix elements
\cite{Birrell:ix,Fulling:nb}. This suggests that {\it analyticity itself is a
unifying principle} for a sensible QFT in analytic, curved spacetime
backgrounds, consistent with various other assumptions but subsuming them.  Much
of the literature on quantum fields in curved spacetime has dealt with free
fields\cite{Mottola:ar,Allen:ux,Birrell:ix,Fulling:nb,Kay:mu}, and has been
concerned with the vexing problem of how to choose the ``right" no-particle
state.  What we are suggesting is that the requirements of constructing a
sensible, {\it interacting} field theory in a curved background may
paradoxically simplify the choice by resolving some if not all of the
ambiguities that may exist for free fields.  

The vacuum for which $n$-point functions obey the requisite analyticity is very
likely unique, since, with Euclidean signature, the differential equations for
propagators become elliptic rather than hyperbolic.  This requirement would also
be consistent with a holographic principle that boundary values uniquely
determine the function everywhere. Such a property seems a desirable starting
point for formulating the conjectured dS/CFT duality\cite{Strominger:2001pn}
and, more generally, for holography in time-dependent backgrounds.  However, the
conjectured implementation of these ideas to date are based on a particular 
choice of MA vacuum\cite{BMS,SV,Strominger:2001pn,Strominger:2001gp}, not on the Euclidean vacuum.

The interpretation of singularities of Feynman diagrams in terms of ``on-shell"
particle properties is well-known in Minkowskian spacetime and it would
obviously be useful to have a generalization to arbitrary spacetime. A
restatement of the Landau rules by Coleman and Norton\cite{C-N} provides a
formulation  amenable to interpretation in coordinate space and potentially
applicable to curved spacetime.  Their result is that ``a Feynman amplitude has
singularities on the physical boundary if and only if the relevant Feynman
diagram can be interpreted as a picture of an energy- and momentum-conserving
process occurring in spacetime, with all internal particles real, on the mass
shell, and moving forward in time."  For the purposes of seeking a similar
theorem in curved spacetime backgrounds, we might restate this by saying that ``a
Feynman diagram has singularities if and only if the internal lines can be
interpreted as classical particles moving on timelike (or null) geodesics between vertices
that are causally related."  Stated in this way, we may conjecture that it is
true for the Euclidean vacuum in an arbitrary, analytic spacetime background
\footnote{The method of proof would have to be rather different from the
familiar ones, relying as they do on properties of amplitudes in momentum
space.}. In Minkowski space, the Landau rules are a reflection of the 
completeness of the particle spectrum, which is the essence of unitarity, 
so the generalization of the Coleman-Norton result to
curved spacetime might be interpreted as an expression of the appropriateness 
of the particle interpretation associated with the Euclidean vacuum.  
Because of the work of Bros and collaborators, reviewed in 
section~\ref{sect:axiomatic}, we are quite confident of the extension of the
Coleman-Norton theorem to the Euclidean vacuum for the de~Sitter background. 
On the other hand, as we described in sections~\ref{sect:treeampl} and
\ref{sect:loopampl}, the MA-vacua have a completely different singularity
structure, requiring a novel, presently unknown, interpretation.  

Assuming that string theory underlies quantum gravity and quantum field theory
in curved spacetime, can one find any motivation for assumptions such as these? 
From its inception in the Veneziano model\cite{veneziano}, a key element in the
development of string theory has been the role of analyticity and the
association of singularities in scattering amplitudes with particles in physical
processes.  It is so much a part of the structure that it is scarcely remarked
upon any more.  It is true that string theory to date can only describe S-matrix
elements and that the relationship of superstrings to nonsupersymmetric theories
is obscure.  Certainly it is not known at this time how to obtain a de~Sitter-like 
background from string theory.  Nevertheless, it may be anticipated that
any effective field theory that comes from string theory will reflect both the
analytic structure of Green's functions familiar from QFT in Minkowski space 
and the association of singularities of Feynman amplitudes with classically 
realizable processes involving particle propagation, as embodied in our 
conjectured generalization of the Coleman-Norton theorem. It certainly 
would be pleasing if this were the case, and it would be even more 
satisfying if analyticity resolved the thorny problem of how to choose 
the correct vacuum state, even if only for a large class of curved backgrounds.  

These considerations clearly will have implications for cosmology in the very
early universe and for physics above the scale relevant to the onset of an
inflationary phase if not beyond the Planck scale.  The popular 
trans-Planckian scenarios\cite{transplanck} that precede the inflationary era
generally employ vacua that are mode-dependent generalizations of the Bogoliubov
transformations (\ref{bogtransf}) leading to the MA-vacua in de~Sitter space.
The simplest construction\cite{danielsson} considers the mode-independent
transformation, equivalent to one particular MA-vacuum; it will therefore be
beset by many of the difficulties emphasized in this paper, such as the
nonthermal character of the background, the noncausal and nonlocal singularities
associated with antipodal points, and the difficulties defining loop diagrams.
The more general constructions will modify the short-distance behavior of 
the MA-propagators without changing their singularities at large distances.  Hence,
they too will confront problems similar to those encountered for the MA-vacua.
Additionally, removing antipodal singularities by modifying the spacetime history
does not remedy the problems within the future lightcone.\footnote{One may 
try to interpret the ``new'' vacua as excitations on the Euclidean vacuum, rather than truly
different vacua; but then one will encounter the well-known problems of infinite
rates of particle production\cite{Fulling:nb}, which will be manifested in
large amounts of energy that does not inflate away.  This energy must be
accounted for at the end of the inflationary era in the transition to a
radiation-dominated cosmology in the usual adiabatic background
\cite{shenker}.}  In summary, our conclusions justify the choice 
of vacuum made in the effective field theory 
description of inflation by Kaloper \etal\cite{kaloper}

In this paper, we have highlighted what we believe to be 
serious challenges to defining and interpreting quantum field theory in
de~Sitter space in a non-Euclidean vacuum.  We suggest that a greater burden of
proof rests on those who would adopt a vacuum in which the propagator is {\it
not} analytic in the usual way.  Their challenge is to show how correlation
functions or observables are to be calculated in a well-defined, unambiguous
manner consistent with QFT and with macroscopic causality. 

\medskip
\centerline{\bf Acknowledgments}

The authors thank D.~Chung for helpful discussions during the initial stages of
this work.  One of us (MBE) offers thanks to H.~ Rubinstein for stimulating his
interest in trans-Planckian scenarios and appreciation to J.~Bros and
U.~Moschella for helpful correspondence concerning their work.  He also thanks
the theory group of LBL for its hospitality, where a portion of this work was
completed.  The other of us (FL) thanks the Aspen Center for Physics for its
hospitality as this work was completed, and J.~Cline, R.~Holman, M.~Kleban,
A.~Lawrence, H.~Ooguri, A.~Rajaraman, and S.~Shenker for stimulating
discussions. This work has been supported in part by the U.S.\ Department of
Energy.

\bigskip
\noindent {\bf Note added:}  As this manuscript was being completed, another
appeared\cite{Banks} that also argued that the MA-vacua are unacceptable.  Our
arguments include problems at the tree level and, at the loop level, the
difficulties we highlight are not specifically tied to the antipodal points.
Secondly, although we have not considered the possibility of identification of
antipodal sector with the causal sector,\cite{verlinde} changing de~Sitter
space to an RP(N) manifold, the problem with loops is already evident in the
mixed $i\epsilon$ prescription associated with the  singularities along the
usual lightcone $Z=1.$  Therefore, we expect to find no alternatives to
choosing the Euclidean vacuum in any case. 


\vfill

\begin{thebibliography}{99}

\bibitem{Mottola:ar}
E.~Mottola,
``Particle creation in de~Sitter space'',
Phys.\ Rev.\ D {\bf 31}, 754 (1985).

\bibitem{Allen:ux}
B.~Allen,
``Vacuum states in de~Sitter space'',
Phys.\ Rev.\ D {\bf 32}, 3136 (1985).

\bibitem{Bunch:yq}
T.~S.~Bunch and P.~C.~Davies,
``Quantum field theory in de~Sitter space: renormalization by point 
splitting'',
Proc.\ Roy.\ Soc.\ Lond.\ A {\bf 360}, 117 (1978).

\bibitem{SSV}
M.~Spradlin, A.~Strominger and A.~Volovich,
``Les Houches lectures on de~Sitter space,''
arXiv:hep-th/0110007.
                              
\bibitem{BMS}
R.~Bousso, A.~Maloney and A.~Strominger,
``Conformal vacua and entropy in de~Sitter space'',
Phys.\ Rev.\ D {\bf 65}, 104039 (2002)
[arXiv:hep-th/0112218].
                         
\bibitem{SV}
M.~Spradlin and A.~Volovich,
``Vacuum states and the S-matrix in dS/CFT'',
Phys.\ Rev.\ D {\bf 65}, 104037 (2002)
[arXiv:hep-th/0112223].

\bibitem{danielsson}
U.~H.~Danielsson,
``A note on inflation and transplanckian physics,''
Phys.\ Rev.\ D {\bf 66}, 023511 (2002)  [arXiv:hep-th/0203198];
``Inflation, holography and the choice of vacuum in de~Sitter space'',
JHEP {\bf 0207}, 040 (2002)  [arXiv:hep-th/0205227].

\bibitem{transplanck}
J.~C.~Niemeyer,
``Inflation with a high frequency cutoff'',
Phys.\ Rev.\ D {\bf 63}, 123502 (2001)
[arXiv:astro-ph/0005533].

R.~Easther, B.~R.~Greene, W.~H.~Kinney and G.~Shiu,
``Inflation as a probe of short distance physics'',
Phys.\ Rev.\ D {\bf 64}, 103502 (2001),
[arXiv:hep-th/0104102].
                             
R.~H.~Brandenberger and J.~Martin,
``On signatures of short distance physics in the cosmic microwave  background'',
[arXiv:hep-th/0202142].
                             
S.~F.~Hassan and M.~S.~Sloth,
``Trans-Planckian effects in inflationary cosmology and the modified  
uncertainty principle'',
[arXiv:hep-th/0204110].

K.~Goldstein and D.~A.~Lowe,
``Initial state effects on the cosmic microwave background and  
trans-planckian physics'',
[arXiv:hep-th/0208167].
                                                     
\bibitem{Balasubramanian:2002zh}
V.~Balasubramanian, J.~de Boer and D.~Minic,
``Exploring de~Sitter space and holography,''
arXiv:hep-th/0207245.

\bibitem{Bros:1990cu}
J.~Bros,
``Complexified de~Sitter space: analytic causal kernels and 
K\"all\'en-Lehmann type representation,''
Nucl.\ Phys.\ Proc.\ Suppl.\  {\bf 18B}, 22 (1991).

\bibitem{Bros:dn}
J.~Bros, U.~Moschella and J.~P.~Gazeau,
``Quantum field theory in the de~Sitter universe,''
Phys.\ Rev.\ Lett.\  {\bf 73}, 1746 (1994).

\bibitem{Bros:1995js}
J.~Bros and U.~Moschella,
``Two-point functions and quantum fields in de~Sitter universe,''
Rev.\ Math.\ Phys.\  {\bf 8}, 327 (1996)
[arXiv:gr-qc/9511019].

\bibitem{Bros:1998ik}
J.~Bros, H.~Epstein and U.~Moschella,
``Analyticity properties and thermal effects for general quantum 
field theory on de~Sitter spacetime,''
Commun.\ Math.\ Phys.\  {\bf 196}, 535 (1998)
[arXiv:gr-qc/9801099].

\bibitem{Tolley:2001gg}
A.~J.~Tolley and N.~Turok,
``Quantization of the massless minimally coupled scalar field and the  dS/CFT
correspondence,''
arXiv:hep-th/0108119.

\bibitem{Witten}
E.~Witten,
``Quantum gravity in de~Sitter space,''
arXiv:hep-th/0106109.

\bibitem{Birrell:ix}
N.~D.~Birrell and P.~C.~Davies,
``Quantum fields in curved space,''
{\it  Cambridge, UK: Univ. Pr. (1982)}.

\bibitem{complcits}
L.~Susskind, L.~Thorlacius and J.~Uglum,
``The stretched horizon and black hole complementarity'',
Phys.\ Rev.\ D {\bf 48}, 3743 (1993)
[arXiv:hep-th/9306069].\newline
L.~Dyson, J.~Lindesay and L.~Susskind,
``Is there really a de~Sitter/CFT duality'',
JHEP {\bf 0208}, 045 (2002)
[arXiv:hep-th/0202163].

\bibitem{coleman}
S.\ Coleman,  
Aspects of symmetry : selected Erice lectures of Sidney Coleman.
{\it Cambridge: Cambridge Univ.\ Press, 1985.}  

\bibitem{Leblond:2002ns}
S.~Gao and R.~M.~Wald,
``Theorems on gravitational time delay and related issues,''
Class.\ Quant.\ Grav.\  {\bf 17}, 4999 (2000)
[arXiv:gr-qc/0007021].\newline
F.~Leblond, D.~Marolf and R.~C.~Myers,
``Tall tales from de~Sitter space. I: Renormalization group flows'',
JHEP {\bf 0206}, 052 (2002)
[arXiv:hep-th/0202094].

\bibitem{ELOP}
See Chapter~2 of R.\ J.\ Eden, \etal,
The Analytic S-Matrix, 
{\it Cambridge: Cambridge Univ.\ Press, 1966.}

\bibitem{Landau}
L.~D.~Landau,
``On analytic properties of vertex parts in quantum field theory'',
Nucl.\ Phys.\  {\bf 13}, 181 (1959).

\bibitem{Cutkosky}
R.~E.~Cutkosky,
``Singularities and discontinuities of Feynman amplitudes'',
J.\ Math.\ Phys.\  {\bf 1}, 429 (1960).

\bibitem{C-N}
S. Coleman and R. Norton, 
"Singularities in the physical region",
Nuovo Cimento {\bf 38}, 438 (1965)

\bibitem{RS}
H.\ Reeh and S.\ Schlieder, 
``Bemerkungen zur Unit\"ar\"aquivalenz von Lorentzinvarianten 
Feldern", Nuovo Cim.\ 22, 1051 (1961), cited in ref.~\cite{SW}.

\bibitem{SW}
R.~F.~Streater and A.~S.~Wightman,
``PCT, spin and statistics, and all that'',
{\it  Redwood City, USA: Addison-Wesley (1989).}

\bibitem{BW}
J.\ J.\ Bisognano and E.\ Wichmann,
J.\ Math.\ Phys.\ 17, 303 (1976).

\bibitem{Strominger:2001gp}
A.~Strominger,
``Inflation and the dS/CFT correspondence'',
JHEP {\bf 0111}, 049 (2001)
[arXiv:hep-th/0110087].

\bibitem{Strominger:2001pn}
A.~Strominger,
``The dS/CFT correspondence'',
JHEP {\bf 0110}, 034 (2001)
[arXiv:hep-th/0106113].

\bibitem{Fulling:nb}
S.~A.~Fulling,
``Aspects of quantum field theory in curved spacetime'',
{\it  Cambridge, UK: Univ. Pr. (1989)}.
                           
\bibitem{shenker}
S.\ Shenker, 
``Inflation as a window into short distance physics",
talk at Strings 2002 Conference, Cambridge, July 15-20, 2002,\newline 
http://www.damtp.cam.ac.uk/strings02/speak.html.\newline                          
N.~Kaloper, M. ~Kleban, A.~Lawrence, S.~Shenker and L.~Susskind,
to appear.

\bibitem{Kay:mu}
B.~S.~Kay and R.~M.~Wald,
``Theorems on the uniqueness and thermal properties of stationary, 
nonsingular, quasifree states on spacetimes with a bifurcate Killing 
horizon'',
Phys.\ Rept.\  {\bf 207}, 49 (1991).

\bibitem{lebellac}
M.~Lebellac, ``Thermal field theory,"
{\it Cambridge: Cambridge Univ. Press, (1996).}

\bibitem{veneziano}
G.~Veneziano,
``Construction of a crossing-symmetric, Regge-behaved amplitude for 
linearly rising trajectories,''
Nuovo Cim.\ A {\bf 57}, 190 (1968).

\bibitem{kos}
P.~Kraus, H.~Ooguri, and S.~Shenker (private communication).

\bibitem{kaloper}
N.~Kaloper, M.~Kleban, A.~E.~Lawrence and S.~Shenker,
``Signatures of short distance physics in the cosmic microwave  
background,''
arXiv:hep-th/0201158.

\bibitem{Banks}
T.~Banks and L.~Mannelli,
``de Sitter Vacua, Renormalization and Locality,''
arXiv:hep-th/0209113.

\bibitem{verlinde}
M.~K.~Parikh, I.~Savonije and E.~Verlinde,
``Elliptic de Sitter Space: dS/Z$_2$,''
arXiv:hep-th/0209120.

\end{thebibliography}
\end{document}